\documentclass[apj]{emulateapj}
\usepackage{natbib,caption2}
\usepackage{epsfig}
\usepackage{amsmath,amssymb,amsfonts,bm}
\usepackage{graphicx}
\usepackage{graphics}
\usepackage{float}
\usepackage{calc}
\usepackage{url}
\usepackage{hyperref}
\usepackage{color}

\submitted{Accepted for publication in {\it The Astrophysical Journal}, 2017 December 6}

\newcommand \beq{\begin{equation}}
\newcommand \eeq{\end{equation}}
\newcommand \bey{\begin{eqnarray}}
\newcommand \eey{\end{eqnarray}}
\newcommand \Myr{\, {\rm Myr} }
\newcommand \Gyr{\, {\rm Gyr} }

\newcommand \pc{\, {\rm pc} }

\newcommand \kpc{\, {\rm kpc} }
\newcommand \msun{M_\odot}

\newcommand \kms{\, {\rm km \, s}^{-1} }
\newcommand \vg{{\bf g}}

\newcommand \dt{{\bf d}t}

\newcommand \divergence{{\bf\nabla\cdot}}

\newcommand \gext{{\bf g}_{\rm ext}}
\newcommand \gint{{\bf g}_{\rm int}}

\newcommand \rp{r_{\rm P}}
\newcommand \rh{r_{\rm h}}
\newcommand \rhf{r_{\rm h,f}}
\newcommand \refe{r_{\rm vir}}
\newcommand \mb{M_{\rm b}}
\newcommand \rhoh{\rho_{\rm h}}
\newcommand \tcr{t_{\rm cr}}
\newcommand \rt{r_{\rm tidal}}

\begin{document}
\title{Gas expulsion in MOND: the possible origin of diffuse globular clusters and ultra-faint dwarf galaxies}
\shorttitle{Gas expulsion in MOND: the possible origin of diffuse GCs and UFDs}
\shortauthors{Wu \& Kroupa}
\author{Xufen Wu\altaffilmark{1,2}, Pavel Kroupa\altaffilmark{3}}\email{xufenwu@ustc.edu.cn}

\altaffiltext{1}{CAS Key Laboratory for Research in Galaxies and Cosmology, Department of Astronomy, University of Science and Technology of China, Hefei, 230026, P.R. China}
\altaffiltext{2}{School of Astronomy and Space Science, University of Science and Technology of China, Hefei 230026, China}
\altaffiltext{3}{Helmholtz-Institut f\"{u}r Strahlen-und Kernphysik, Universit\"{a}t Bonn, Nussallee 14-16, D-53115 Bonn, Germany}



\begin{abstract}
  We study the evolution of star clusters located in the outer regions of a galaxy undergoing a sudden mass loss through gas expulsion in the framework of Milgromian dynamics (MOND) by means of N-body simulations. We find that, to leave a bound star cluster, the star formation efficiency (SFE) of an embedded cluster dominated by deep MOND gravity can be reduced down to $2.5\%$. For a given SFE, the star clusters that survive in MOND can bind a larger fraction of mass compared to the Newtonian dynamics. Moreover, the more diffuse the embedded cluster is, the less substantial the size expansion of the final star cluster is. The density profiles of a surviving star cluster are more cuspy in the centre for more massive embedded clusters, and the central density profiles are flatter for less massive embedded clusters or for lower SFE. This work may help to understand the low concentration and extension of the distant low-density globular clusters (GCs) and ultra-faint and diffuse satellite galaxies around the Milky Way.

\end{abstract}

\section {Introduction}
Star clusters form in dense cloud clumps in giant molecular clouds (GMCs) and most stars form in star clusters \citep{Kroupa1995a,Lada_Lada2003,Gieles+2012,Belloni+2017}. After the formation of OB stars, the remaining gas in a star cluster is expelled on a time scale of less than a few $\Myr$ through multiple mechanisms including stellar winds, Type II supernovea explosions, ionizing radiation, radiation pressure \citep{Hopkins+2013,Dib+2013,Krumholz_Matzner2009,Lada2010,Murray+2010}. The dense clumps are disrupted and star formation is terminated. The gravitational potential of the star cluster becomes shallower after the sudden removal of gas, and therefore stars escape from the star cluster. Moreover, the star cluster subsequently expands \citep[e.g.,][]{Goodwin1997,Kroupa+2001,Bastian_Goodwin2006,Goodwin_Bastian2006,Boily_Kroupa2003,Baumgardt_Kroupa2007,Marks+2012,Pfalzner_Kaczmarek2013,Brinkmann+2017,Banerjee_Kroupa2017}. 

GCs are $10-12~\Gyr$ old self-bound systems with a low metalicity and a negligible of gas mass. Therefore the star formation must be finished already during an early stage of the GCs' life and the residual gas must be blown out of the GCs efficiently to prevent a new generation of star formation from the metal-rich gas \citep{Goodwin1997,Goodwin_Bastian2006,Bekki+2017}. In the abrupt process of gas expulsion driven by the supernovae or an accumulative effect of ionisation and stellar winds from the first generation of massive stars, the star clusters may be completely destroyed since the escape of stars again drives the gravitational potential of the star cluster to become shallower \citep{Boily_Kroupa2003}.

In early observations, the star formation efficiency (SFE) for the overall GMCs in global measurements in both Milky Way and external galaxies was found to be $\approx 2\%$ \citep{Evans1991,Kennicutt1998}. Recent observations on a large sample of GMCs in the Milky Way by \citet{Murray2011} showed that the SFE for the GMCs are in the range $[0.2\%,~20\%]$, and a luminous-weighted average is $8\%$. Although the SFEs are larger in the dense cloud clumps of molecular clouds \citep{Lada_Lada2003}, it has been confirmed by extensive observations that the SFE for the dense cloud clumps of molecular clouds is no more than $30\%$. Subsequent observations revealed that the SFE for a sample of dense cloud clumps in GMCs falls in the range of $1.2\%-22\%$, with a mean of about $10\%$ \citep{Higuchi+2009,Kainulainen+2014,Megeath+2016}. The value of SFE has been extensively studied within the framework of standard Newtonian dynamics. A critical value of the SFE of $33\%$ is obtained for a single and initially virialised star cluster to leave a bound remnant \citep{Goodwin1997,Boily_Kroupa2003,Baumgardt_Kroupa2007}. For star clusters that form in complexes and in a merging process within the gas expulsion time-scale, the SFE can be as low as $20\%$, which is able to leave a bound remnant from the voilent gas expulsion \citep{Fellhauer_Kroupa2005}, which may explain the formation of the faint GCs in the outer Galactic halo \citep{Harris1996}. 

For a star cluster to survive gas expulsion with an ultra-low SFE in the dense clumps of GMCs \citep[such as][]{Higuchi+2009}, the gravitational potential needs to be deeper than that generated by Newtonian dynamics. In principle, dark matter could potentially provide such a deep potential. However, it has been well established that dark matter does not exist in GCs \citep{Moore1996}. It is therefore necessary to invoke a different mechanism. Here we propose that Milgromian Dynamics \citep[known as modified Newtonian dynamics or MOND,][]{Milgrom1983a,BM1984} can naturally provide a sufficiently deep potential. Comparing to Newtonian dynamics in which the potential increases as $1/r$ at large radii of a system, the potential increases logarithmically for large $r$ within the framework of MOND. In the process of sudden gas expulsion, a star cluster formed in a dense cloud clump undergoes an abrupt mass loss, and the dynamics of the system rapidly transits into deep MOND limit. For a given SFE, the self-gravity of the stellar component is stronger in MOND than that in Newtonian gravity. As a result, the system can bind more high energy stars, which indicates that MOND may allow a lower SFE in an embedded cluster \footnote{In the present models an embedded cluster is defined to be the star cluster with all its stars having formed and at the time before the unused gas is expelled. This is a theoretical description which encompasses the main relevant physical process in that the potential of the forming cluster is dominated by the gas until a time when most of the star formation has ceased and the gas is blown out. In reality the evolution from a pure gaseous proto-cluster to the exposed very young cluster through the embedded phase is gradual and takes about 1Myr in total, while gas-removal after 0.6~Myr has been found to account for observed exposed clusters very well \citep{Kroupa+2001,Banerjee_Kroupa2012,Banerjee_Kroupa2015,Banerjee_Kroupa2017}. Assuming a spherically symmetric smooth embedded cluster model is a physically plausible initial state, because this work on young clusters has also shown that the merging of many sub-clusters takes too long to account for the observed density and velocity dispersion profiles for the ages ($1-3~\Myr$) of the observed exposed clusters. The model thus rests on the physically plausible description that individual stars take about 0.1~Myr to assemble about $95~\%$ of their final mass, independently of their mass \citep{Duarte-Cabral+2013,Wuchterl_Tscharnuter2003,Motte+2017}, and that at this stage the star decouples from the hydrodynamics and becomes a ballistic particle orbiting within the virialising potential. The cluster takes about $1~\Myr$ to form its stars, such that, when the gas is expelled, it can be assumed to be close to virial equilibrium.}
leaving a self-gravitationally bound remnant. Moreover, since a larger mass is bound to the final GC for the given SFE, the size of the GC is expected to be larger in MOND for a compact embedded cluster. In the case of a diffuse embedded cluster, the MOND potential is much deeper and the size of the final GC should be smaller than that in Newtonian dynamics.
 Accordingly, a more extended and more diffuse star cluster can survive in MOND, but cannot in Newtonian dynamics. As far as we know, none of such examinations have ever been performed.

In this paper, we study the sudden gas expulsion process for star clusters with ultra-low SFE in MOND and examine whether there is a bound remnant. The sudden gas expulsion of star clusters with a SFE in the range $[2.5\%, ~50\%]$ in MOND is considered. N-body initial conditions (ICs) for the initially compact and spherically symmetric models are constructed for the embedded clusters in Sec. \ref{ics}. The sudden gas expulsion process is studied by applying a sudden decrement of mass according to the values of SFE for the isolated embedded clusters in Sec. \ref{gas}. The models are evolved for $1~\Gyr$ using MOND N-body simulations. The fraction of bound mass to the ICs, the evolution of Lagrangian radii and the normalised density profiles are studied in Sec. \ref{bound}. Finally, the results are summarised and discussed in Sec. \ref{summary}.


\section{Models of gas expulsion}
\subsection{ICs for embedded clusters in MOND}\label{ics}
MOND has been tested for decades. On large scales, MOND encounters several unsolved challenges. The strong and weak lensing data from clusters of galaxies showed that MOND also requires dark matter such as neutrinos \citep{Natarajan_Zhao2008,Angus+2007}, and one of the most famous example is the Bullet Cluster 1E0657-56 \citep{Clowe+2006}. Although there are several relativstic versions of MOND, including Bekenstein's Tensor-Vector-Scalar theory \citep[TeVeS,][]{Bekenstein2004} and Milgrom's BIMOND \citep{Milgrom2009b,Milgrom2010}, claiming that the convergence mape of the cluster can be reproduced without any dark matter, these theories bring in new problems. The convergence map of the Bullet Cluster remains a major challenge for MOND so far. Moreover, the ring-like dark matter structure around the rich galaxy cluster Cl0024+17 is also hard to understand within the framework of MOND \citep{Jee+2012}. Besides the difficults on the scale of clusters of galaxies, the Cosmic Microwave Background (CMB) radiation is another problem for MOND. To be consistent with the observations of the third peak of the acoustic power specturm of the CMB, neutrinos as dark matter are required \citep{Skordis+2006,Angus2009,Zhao2008}. Moreover, even in the presence of neutrinos, MOND is difficult to form the correct amplitude for the mass function of galaxy clusters \citep{Angus_Diaferio2011,Angus+2013,Angus+2014b}.

Although MOND faces the challenges on large scales, it is very successful and promising on scales of galacxies and star clusters. It can successfully predict the rotation curves for all galaxies, including our Milky Way Galaxy \citep[e.g.,][]{Sanders_McGaugh2002,Famaey_McGaugh2012,Famaey_Binney2005,Famaey+2007}. Moreover, MOND provides a unified explanation on the escape velocity of galaxies \citep{Wu+2007,Wu+2008,Banik_Zhao2018}, the rotational speed in polar ring galaxies \citep{Lughausen+2013}, the formation of the shell structure in NGC 3923 \citep{Bilek+2013,Bilek+2014}, the velocity dispersion of M31 dwarf galaxies \citep{McGaugh_Milgrom2013,McGaugh2016}, the mass discrepancy-acceleration correlation of disc galaxies \citep{Milgrom1983b,Sanders1990,McGaugh2004,Wu_Kroupa2015} and of pressure-supported galaxies \citep{Scarpa2006}， and the relation between the baryonic and dynamical central surface densities for disc galaxies \citep{Milgrom2016}. 

Based on a critical test of MOND developed in Bonn \citep{Baumgardt+2005}, Newtonian dynamics can explain well the kinematics and dynamics of some distant GCs, including NGC 2419 \citep{Baumgardt+2009,Ibata+2011a,Ibata+2011b}, Palomar 4 \citep{Frank+2012} and Palomar 14 \citep{Jordi+2009}. These star clusters behave Newtonian with small values of the velocity dispersion. However, \citet{Gentile+2010} argued that using a small sample of stellar kinematics data (17 stars) for Pal 14, it is insufficient to falsify MOND. \citet{Sanders2012a,Sanders2012b} showed that polytropic models in MOND for NGC 2419 can fit well the observations of the surface brightness and the velocity dispersion profiles, and therefore NGC 2419 might not be a problem for MOND. On the other hand, recently, a number of observations and simulations on Galactic GCs suggest departures from Newtonian gravitation. In these studies, the line-of-sight (LoS) velocity dispersion profiles of such GCs, $\sigma_{\rm LoS}(r)$, are flat at large radii \citep{Scarpa+2007,Scarpa+2011,Scarpa_Falomo2010,Lane+2009,Lane+2010,Hernandez_Jimenez2012,Hernandez+2013,Durazo+2017}. The observed flat $\sigma_{\rm LoS}(r)$ profiles are apparently at odds with Newtonian dynamics, which predicts that $\sigma_{\rm LoS}(r)\propto r^{-1/2}$ at large radii, but MOND can naturally reproduce the observed outer $\sigma_{\rm LoS}(r)$ of the distant GCs \citep{Milgrom1994}.

It turns out that MOND gives a good discription of the kinematics and dynamics on the scales of GCs and also the central \citep{Milgrom2009} and outer regimes (\citealt{Kroupa+2012,Kroupa2012,Kroupa2015} and also see the review of \citealt{Famaey_McGaugh2012}) of galaxies.

\begin{table}
\begin{center}\vskip -0.0cm
  \caption{Parameters of ICs for the embedded clusters with a Plummer density profile: the $1_{\rm st}-5_{\rm th}$ columns are: ID of the ICs, the total mass $M$, the Plummer radius $\rp$, the half-mass radius $\rh$ and the central density $\rhoh \equiv \frac{M_{\rm b}}{2} \rh^{-3}$. The sixth column is the $3-$dimensional velocity dispersion, $\sigma$, of the overall system. The $8_{\rm th}-9_{\rm th}$ columns are the the radius enclosing $90\%$ of mass, $r_{\rm 90}$, and the crossing time, $\tcr \equiv r_{\rm 90} / \sigma$. We refer to initial models with larger $\rp$ and $\rh$ as being initially more diffuse.}
\begin{tabular}{lcccccccc}
\hline
ICs& $M$ & $\rp$ & $\rh$ &  $\rhoh$& $\sigma$ & $r_{\rm 90}$& $\tcr$  \\
\hline
& $\msun$ & $\pc$& $\pc$ &  $\msun \pc^{-3}$ & $\kms$ & $\pc$& $\Myr$  \\
\hline
1 & $10^7$ &5.0& 6.5 & $1.80\times10^4$ & 50.7 & 18.5 & 0.4 \\
2 & $10^6$ &5.0& 6.5 & $1.80\times10^3$ & 16.8 & 18.5& 1.1\\
3 & $10^5$ &5.0& 6.5 & $1.80\times10^2$ & 6.5 & 18.5 &2.8 \\
4 & $10^5$ &10.0&13.1& $2.25\times10^1$ & 5.8 & 36.7 & 6.3\\
\hline
\end{tabular}
\label{plummer}
\end{center}
\end{table}

The MOND Poisson equation that satisfies the conservation laws of energy, momentum and angular momentum is \citep{BM1984}
\beq\label{mond}
\divergence [\mu(X) { \vg}]=4\pi G\rho_{\rm b},  \,\,\,\, 
X=|{\vg}|/a_0 \, ,
\eeq
where ${\vg}$ is the gravitational acceleration in MOND, and $a_0=3.7\pc\Myr^{-2}$ is Milgrom's gravitational constant. The interpolating function $\mu\rightarrow 1$ when $X \gg 1$ and $\mu \rightarrow X$ when $X \ll 1$, corresponding to the Newtonian and MONDian limits, respectively. In the deep MOND limit, the gravititaional acceleration $g=\sqrt{a_0 g_{\rm N}}$, where $g_{\rm N}$ is the Newtonian gravity acceleration. The circular velocity, $v_{\rm c}$, following from the centrifugal acceleration, is
  \beq\label{vcirc} v_{\rm c} = (GMa_0)^{1/4},\eeq
where $M$ is the baryonic mass of the system. Eq. \ref{vcirc} implies that a baryonic system is embedded in a logarithmic phantom dark matter halo potential, when interpreted in terms of Newtonian dynamics. This is an effective dark matter halo. The mass of the baryonic matter together with the phantom dark matter halo is the Newtonian dynamical mass of the system.

In what follows, a simple form of the $\mu$ function \citep{Famaey_Binney2005} will be used,
\beq
\mu(X)=\frac{X}{1+X},
\eeq
which fits better the terminal velocity of the Milky Way and NGC 3198. The simple $\mu$ function transits a system from the deep MOND limit to the Newtonian limit more gradually than the 'standard' $\mu$ function of \citet{Milgrom1983b}, and works better in both very weak and very strong gravities \citep{Zhao_Famaey2006}.

Before the gas expulsion, the embedded clusters are more compact and more massive than the present-day GCs. \citet{Plummer1911}'s profile is chosen for the density distribution of embedded clusters as follow,
\beq
\rho_{\rm b} (r)=\left(\frac{3\mb}{4\pi \rp^3} \right)\left(1+\frac{r^2}{\rp^2}\right)^{-5/2},
\eeq
where $\mb$ and $\rp$ are the total mass and the scale length of the Plummer model. The stellar and gaseous density profiles are assumed to follow the Plummer profile with the same Plummer radius, an assumption which has proven to lead to successful modelling of the Orion Nebula Cluster and the Pleiades \citep{Kroupa+2001}, of NGC 3603 \citep{Banerjee_Kroupa2015}, and of R136 \citep{Banerjee_Kroupa2012} and is physically plausible in that the local star formation rate is approximately propotional to the local gas density.

The MONDian potentials of the Plummer models are calculated using a MOND Possion solver \citep{Nipoti+2007,nmody}. The N-body ICs with an isotropic velocity dispersion in MOND gravity are constructed using Lucy's method \citep{Lucy1974}, and the code was originally implemented by \citet{Gerhard1991} for constructing both isotropic and anisotropic N-body models in Newtonian gravity. Here we simply replace the Newtonian potential and circular velocity with the MONDian potential and circular velocity in the N-body generator. A series of self-consistent isotropic N-body initial conditions (ICs) are constructed thereby. The parameters of the models are summerised in Table \ref{plummer}. Note that our ICs for the embedded clusters are in equilibrium, i.e., the virial ratio, $Q_0 \equiv \frac{T_0}{|W_0|} =0.5$, where $T_0$ and $W_0$ are the initial kinetic energy and initial potential energy for the models. This is a physically plausible assumption because stars form in the cloud core in a few crossing times and not instantly at the same time, and the bulk of the forming embedded cluster will be close to the $Q_0=0.5$ state when the gas expulsion occurs. There are $100,000$ equal-mass particles in each model.

\subsection{N-body realisation for the gas expulsion}\label{gas}
Since MOND introduces a larger dynamical mass compared to a pure Newtonian system with the same density distribution, especially for diffuse systems, a MOND gravitational potential can bind more stars with high energy. Therefore, a lower value of SFE is allowed in MOND. To examine this MOND effect, we assume that the values of the SFE range from $10\%$ to $50\%$ with a $10\%$ interval of increment. Further, the SFE is reduced down to $5\%$ and $2.5\%$ for all the embedded cluster models. Star clusters cannot survive in Newtonian gravity with such small values of the SFE in any of the existing studies \citep{Lada+1984,Goodwin1997,Boily_Kroupa2003,Fellhauer_Kroupa2005,Baumgardt_Kroupa2007,Shukirgaliyev+2017}.
Because the models are massive and the number of particles is large, two-body relaxation can be ignored in the simulations.  It is therefore sufficient to simulate the gas expulsion by means of a collisionless N-body code. The orbits of the N-body systems can by integrated by using the particle-mesh N-body code \emph{NMODY} \citep{nmody}, which solves gravitational accelerations and potentials in both standard Newtonian and Milgromian dynamics. More details and tests on the code in both dynamics can be found in \citet{Nipoti+2007,Nipoti+2008,Nipoti+2011,Wu+2013} and \citet{Wu+2017}. In the following simulations, we use a grid-resolution of $n_r\times n_{\theta} \times n_{\phi}=256 \times 32 \times 64$, where $n_r,~n_\theta,~n_\phi$ are the number of grid cells in radial, polar and arthimuthal dimensions. The radial grids are segmented by $r_i = r_s\times \tan \left[(i+0.5)0.5\pi /(n_r+1)\right]$ with $r_s=20\pc$ and $i=0,1,2,...,n_r$, the angular grids are equally segmented, and the angular resolution of the spherical harmonic expansion for the Poisson solver at each time step is $l_{\max}=8$.

\begin{figure}
\includegraphics[width=90mm]{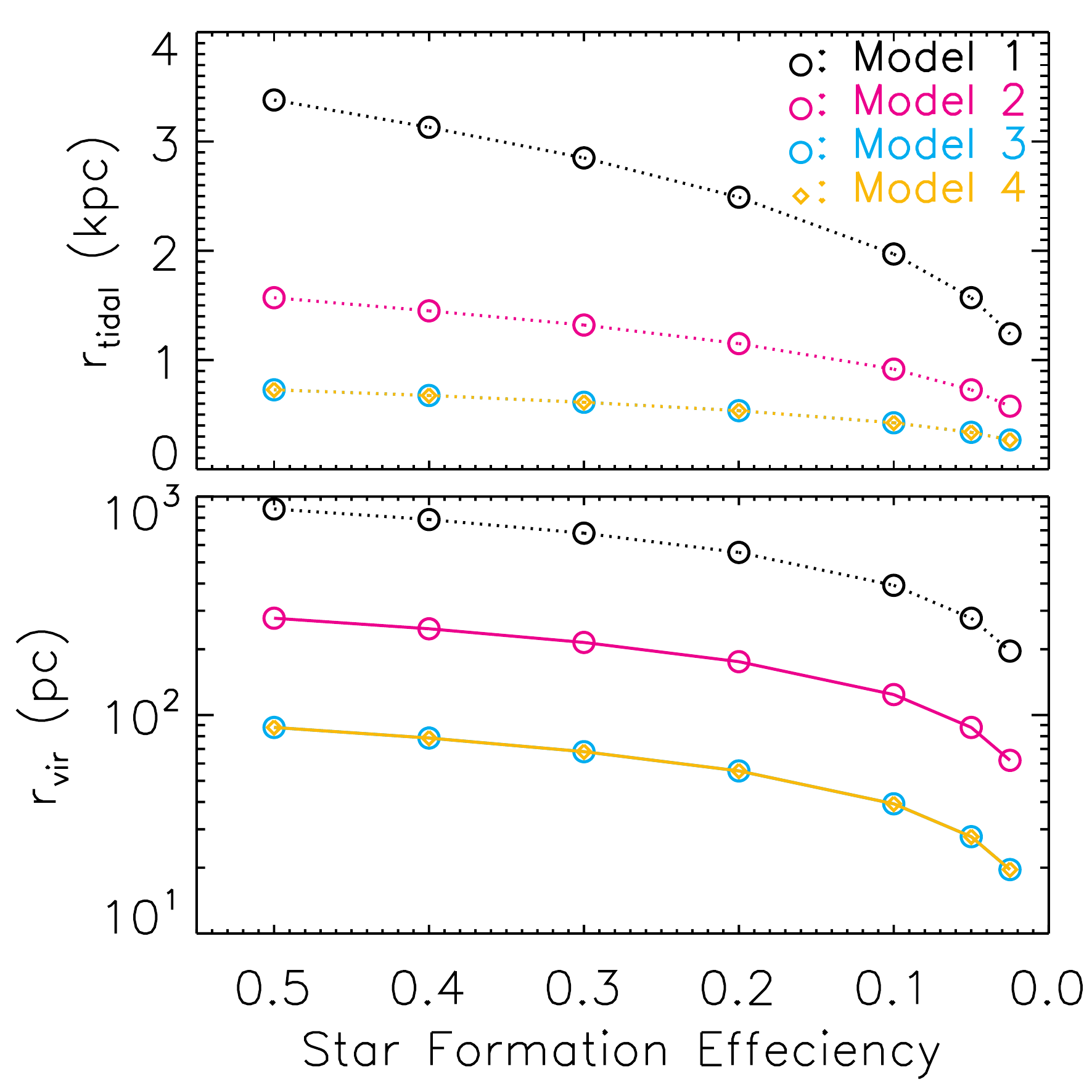}
\caption{Upper panel: the values of tidal radius, $\rt$, of the initial star clusters assuming a Galactocentric distance of $100~\kpc$. Lower panel: the values of virial radii, $\refe$, of the initial embedded star cluster. }
\label{rtidal}
\end{figure}

The $3$-dimensional velocity dispersion, $\sigma$, of the overall embedded clusters including gas and stars are calculated and shown in Table \ref{plummer}. The crossing time of the embedded clusters is defined as $\tcr \equiv r_{\rm 90} /\sigma$, where $r_{\rm 90}$ is the $90\%$ Lagrangian radius. Note that the crossing time in MOND is shorter than that in a Newtonian model with the same baryonic mass density distribution, because the value of $\sigma$ is larger in the deeper MOND potential.

The sudden gas expulsion is modelled by changing the fraction of mass for all particles immediately, i.e., the initial particle-mass multiplies the SFE in each simulation. It has been shown that a more gradual removal of gas will leave a bound remnant with a lower SFE \citep{Baumgardt_Kroupa2007}. We will focus on the most extreme case of gas expulsion in MOND. The effect of gradual gas expulsion or the effect of a different initial stellar mass function is beyond the scope of this paper, and will be studied in a future project. The global time steps are defined as $\dt= \frac{0.1}{\sqrt{\max |\nabla \cdot {\bf g}|}}$, and thus there are about 10 time steps for a circular orbit in the densest regime. We freely evolve the GCs in their new post-gas expulsion self-gravity for about $1~\Gyr$. This is over $150~\tcr$ for the most diffuse initial embedded cluster model (model $4$) and is over $300~\tcr$ for the other models.

\subsection{Truncation radius for the final products}
For an isolated protocluster model in MOND, ideally, no stars can escape from the logarithmic potential. However, the external field truncates the logarithmic potential well to $1/r$ dependency at large radii and enables stars to escape \citep{Famaey+2007,Wu+2007}. The external field defines the tidal radius of a star cluster in a host galaxy \citep{Zhao_Tian2006},
\bey
&\rt&=\left( \frac{M_{\rm b}^{\rm GC}}{(1+\eta)M_{\rm gal}} \right)^{1/3}D_0, \\
&\eta&=-\frac{d\ln g}{d\ln D_0},\nonumber
\eey
where $\eta \rightarrow 2$ in the Newtonian limit and $\eta \rightarrow 1$ in the deep MOND limit. $M_{\rm b}^{\rm GC}$ and $M_{\rm gal}$ are the baryonic mass of the embedded star cluster (i.e., the mass of the stellar component in the embedded cluster before the gas expulsion. Note that here the mass of the gas component is not included.) and of the galaxy, respectively. $D_0$ is the distance between the young star cluster and the centre of the host galaxy. The baryonic Besan{\c c}on Milky Way model \citep{Wu+2007} is used here as the host galaxy, with $M_{\rm gal} \approx 6.5\times 10^{10}\msun$. A Galactocentric distance of $100~\kpc$ is assummed for the star clusters, which stands for the distance of remote GCs such as AM 1 ($123~\kpc$), Pal 3 ($96~\kpc$), Pal 4 ($112~\kpc$), NGC 2419 ($89~\kpc$, \citealt{Harris1996}). 
The values of the tidal radii for the initially embedded star clusters as a function of SFE are shown in the upper panel of Fig. \ref{rtidal}. The tidal radius of a star cluster in MOND is larger than that in Newtonian gravity, since $M_{\rm gal}$ in Newtonian dynamics should include both baryonic and dark matter, which is much larger than the mass of pure baryonic matter in the MONDian Galaxy, and because the cluster generates a phantom dark matter halo around itself, boosting its effective Newtonian mass. 

Besides the tidal truncation of the star clusters, the uniform gravitational background field from the Galaxy plays an important role for the star clusters. A self-bound system in Newtonian dynamics is not affected by such a uniform external field, and this is known as the Strong Equivalence Principle (SEP), which is one of the basic assumptions of Einstein's theory of general relativity. However, the SEP is violated in MOND \citep{Milgrom1983a,BM1984}. Possible evidence for SEP violation has been found by \citet{Wu+2010,Wu+2017} and \citet{Thomas+2017}. The dynamics of a self-bound system is governed by both the internal and the external gravitational fields, i.e., ${\bf g} = \gint + \gext$ in Eq. \ref{mond}. A truncation radius can be roughly estimated in MOND within an external field,
\beq
\refe = \sqrt{GM_{\rm b}^{\rm GC}a_0}/g_{\rm ext},
\eeq
where the strength of the internal gravity equals that of the external field. At a radius larger than $\refe$, the system is dominated by the external field. $\refe$ is the virial radius for a self-bound system \citep{Wu_Kroupa2015}. The phantom dark matter halo is truncated at $\refe$, and the mass of the phantom dark matter halo, sourced purely by the stars (the gas component is not used here as it is removed rapidly), is
  \beq M_{\rm phantom} = M_{\rm b}^{\rm GC}a_0/g_{\rm ext}. \eeq
At a Galactocentric distance of $100~\kpc$, the strength of the external field from the Galaxy is $g_{\rm ext}\approx 0.087a_0$. Since the external field from the Milky Way is weak, the star clusters are simulated in isolation. The external field effect mainly reflects the truncation radii for the star clusters.

The virial radii of the embedded star clusters are displayed in the lower panel of Fig. \ref{rtidal}. The virial radii are smaller than the tidal radii for all the models, indicating that the uniform external field dominates the dynamics in the star clusters at a smaller radius than the tidal field. Therefore, the external field truncates a star cluster at $\refe$. At the radius where $r>\refe$, the rotation curve of a system behaves Newtonian-like, but with the velocities being a factor $\frac{1}{\mu_{\rm ext}}=\frac{a_0}{|g_{\rm ext}|}$ larger within the weak external field.


\section{Results}\label{bound}
\subsection{The loss of mass}\label{mloss}
\begin{figure}
\includegraphics[width=90mm]{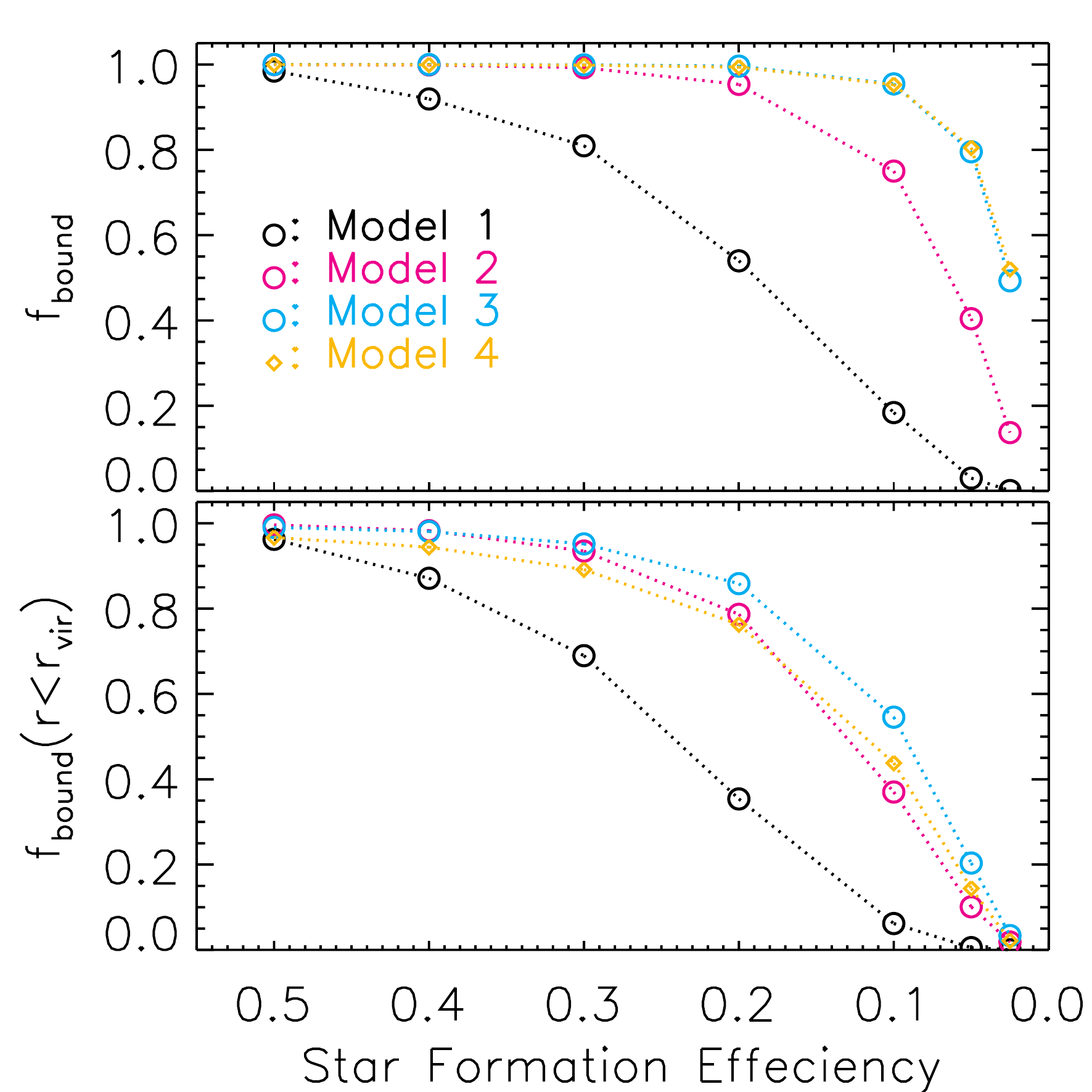}
\caption{Bound mass fraction as a function of the SFE in MOND. $f_{\rm bound}$ is calculated assuming a truncation by $\rt$ in the upper panel, and a truncation by $\refe$ in the lower panel.}
\label{fbound}
\end{figure}

The bound mass fraction at the end of the simulations, $f_{\rm bound}$, as a function of the SFE in MOND is shown in Fig. \ref{fbound}. The bound particles are defined as stars with binding energy $E_{\rm bind}=T+W < 0$ within the initial truncation radius of the stellar component in the embedded cluster shown in Fig. \ref{rtidal}. $f_{\rm bound}$ is the ratio between the bound stars at the end of the simulation and the initial stellar mass in the embedded cluster. The upper panel shows the fraction of bound mass truncated by the tidal radius, and the lower panel presents the fraction of bound mass truncated by the external field. 

In general, for a given model with a fixed SFE, the fraction of bound mass truncated by the external field is smaller than that by the tidal field. Since both effects should be taken into account in a MOND system, we shall truncate the star cluster at $\refe$, which is much smaller than $r_{\rm tidal}$ (see Fig. \ref{rtidal}). The values of $f_{\rm bound}$ for the surviving star clusters truncated at $\refe$ are also listed in Table \ref{tab-fbound}. The first row of the table indicates the values of the SFE and the $2_{\rm nd}-5_{\rm th}$ rows show $f_{\rm bound}$ for different values of the SFE.

In models with larger values of SFE, i.e., $40\%$ and $50\%$, self-bound cores can survive after the gas expulsion in Newtonian dynamics. Thus it is possible to compare the models, with the same density profiles and the same SFE, defined in MOND and in Newtonian dynamics. Here, we find that all the MOND models leave the majority, i.e. over $85\%$, of mass being self-bound after the gas is expelled immediately. In the previous studies in Newtonian dynamics, e.g., in \citet{Boily_Kroupa2003}, $f_{\rm bound}$ is only $66\%$ for a SFE of $50\%$, and in \citet{Baumgardt_Kroupa2007}, $11\%-34\%$ and $13\%-72\%$ of mass remains bound at the end of their simulations corresponding to a SFE of $40\%$ and $50\%$, respectively.  The reasons are: i) the tidal radii at the same Galactocentric distance are much smaller in Newtonian dynamics; ii) the external field effect in MOND dominates the dynamics of the star cluster only in the outer regime where $r>\refe$. Therefore, with the same SFE, a larger fraction of stars can be bound to the star cluster after sudden gas expulsion.

With a low value of SFE, $10\%$, all the MOND models can leave a bound core. $f_{\rm bound}$ decreases with the reduction of the SFE, especially for the most massive initial model, model $1$. Approximately, $6\%$ of the stellar component is bound at the end of the simulation with a SFE$=10\%$, and about $37\%$ stars remain bound with a SFE$=20\%$. The less massive models (i.e., models $2$-$4$) have larger values of $f_{\rm bound}$, since these models are more dominated by deep MOND gravity, and their potentials are significantly deeper than in Newtonian dynamics. Remarkably, for models $2$, $3$ and $4$, the values of $f_{\rm bound} \approx 0.37,~0.55$ and $0.44$, respectively, for a SFE of $10\%$. For the three models, $f_{\rm bound}(r<\refe)>75\%$ with SFE$ \ge 20\%$. This implies that most of the stars are bound to the originally diffuse models after gas expulsion. This is never expected in a Newtonian model, in which GCs are completly destroyed for such an abrupt gas expulsion event with a SFE lower than $33\%$ \citep{Goodwin1997,Boily_Kroupa2003,Baumgardt_Kroupa2007}. Although it is possible for a star cluster to form a bound core with a lower value of the SFE in Newtonian gravity, for instance, for adiabatic gas expulsion \citep{Baumgardt_Kroupa2007}, an initially dynamically cold star cluster \citep{Goodwin2009} and star clusters forming in a deeper potential of a star-cluster complex \citep{Fellhauer_Kroupa2005,Smith+2011}, these physical effects should also be valid in MOND, i.e., increase the survivality of star clusters. We shall not introduce such physics into this work, given that a single star cluster formed in equilibrium undergoing an immediate gas expulsion presents an evident and robust difference between MOND and Newtonian dynamics.

Moreover, simulations of the process of gas expulsion with a SFE of $5\%$ and $2.5\%$ are performed here. The results of bound mass in these simulations are presented in Fig. \ref{fbound} as well. Remarkably, it is possible to leave a bound core with a SFE of $2.5\%$ for models $2$-$4$, and $f_{\rm bound}(r<\refe) \approx 2\%-3\%$. The bound fraction is over $10\%$ for the three models with SFE=$5\%$.

Note that two-body relaxation is ignored in the simulations. For the least massive models, models $3$ and $4$, with a SFE of $2.5\%$, the mass of the embedded clusters are only $2500~\msun$, and the bound fraction of the stars is only a few percent. Therefore two-body relaxation could play an important role in these ultra-low SFE systems. The bound core might not be able to survive for $1~\Gyr$ considering the effect of two-body relaxation. For models with a SFE of $5\%$, the bound fraction of stars is one order of magnitude larger than that with a SFE of $2.5\%$, and the two-body relaxation effects can be safely ignored. The effect of two-body relaxation for low mass systems with low SFE in MOND will need to be studied in future work.

\begin{table}
\begin{center}\vskip -0.0cm
  \caption{The fraction of bound mass, $f_{\rm bound}$, of the surviving star clusters after gas expulsion evaluated at an age of $1~\Gyr$. The star clusters are truncated at $\refe$ by the external field effect. Columns $2-8$ show $f_{\rm bound}$ for different values of the SFE. ICs are tabulated in Table \ref{plummer}.}
\begin{tabular}{lcccccccc}
\hline
ICs& SFE$=0.5$ & $0.4$ & $0.3$ &  $0.2$& $0.1$ & $0.05$& $0.025$  \\
\hline
1& $f_{\rm bound}=$0.96 & 0.87 & 0.69 & 0.35 & 0.06 & 0.01 & 0.000 \\
2& 1.00 & 0.98 & 0.93 & 0.79 & 0.37 & 0.10 & 0.018 \\
3& 0.99 & 0.98 & 0.95 & 0.86 & 0.55 & 0.20 & 0.034 \\
4& 0.97 & 0.94 & 0.89 & 0.76 & 0.44 & 0.14 & 0.022 \\
\hline
\end{tabular}
\label{tab-fbound}
\end{center}
\end{table}

\subsection{Lagrangian radii and half mass radius}\label{rmass}

\begin{figure*}
\begin{centering}\includegraphics[width=140mm]{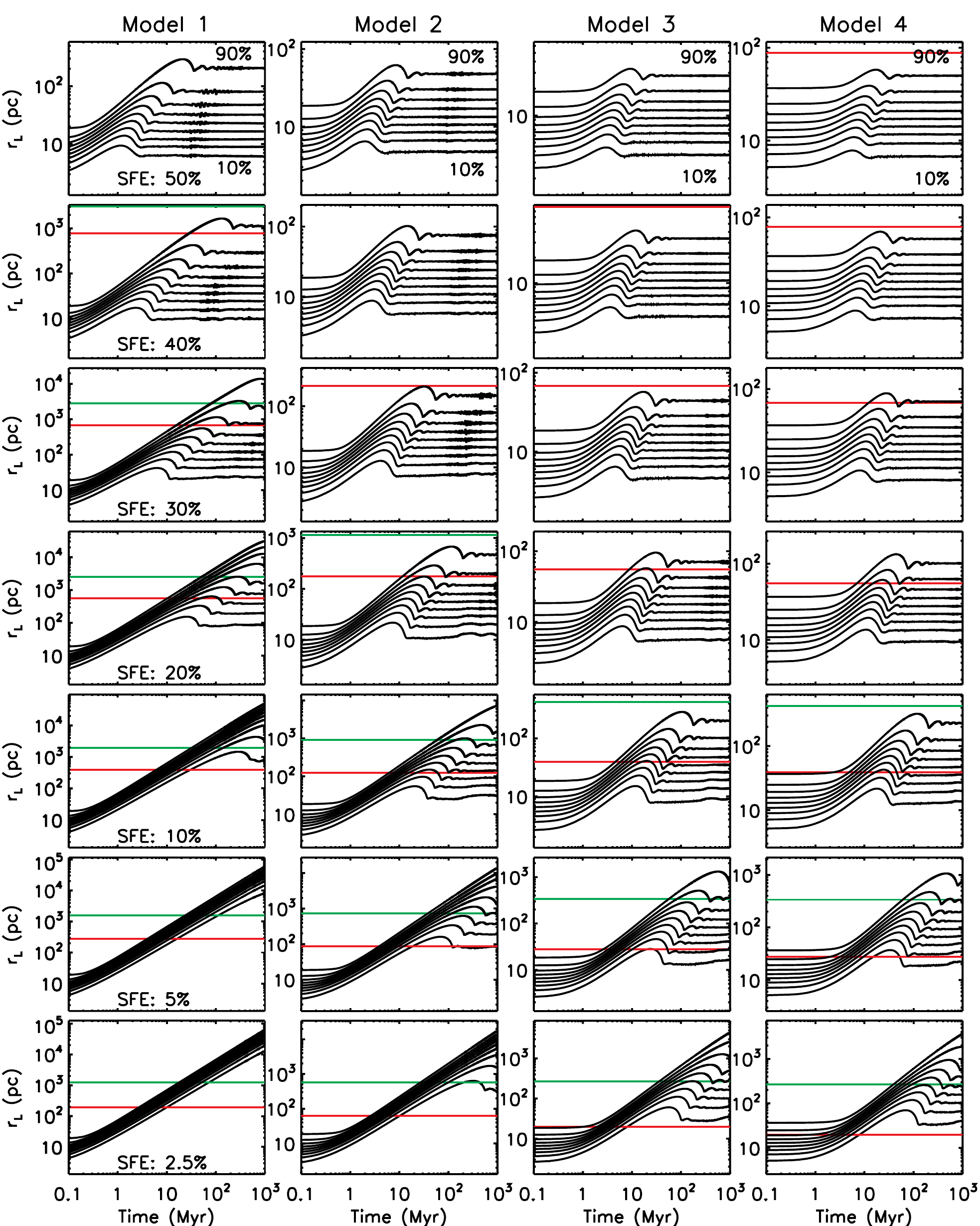}
  \caption{Lagrangian radii (labelled as $r_{\rm L}$ in the vertical axes) increasing from $10\%$ to $90\%$ in steps of $10\%$, $L_{0.1},~...,~L_{0.9}$, of the star clusters undergoing sudden gas expulsion. The panels in columns from left to right present models $1-4$, and the panels in rows from top to bottom show the results for models with the SFE from $50\%$ to $2.5\%$. The green and red lines indicate the initial tidal radius, $\rt$, and the initial external field truncation radius, $\refe$, respectively.}
  \end{centering}
\label{lr}
\end{figure*}

Fig. \ref{lr} illustrates the evolution of Lagrangian radii, i.e., the spherical radii enclosing different fractions of the initial stellar mass in a model, increasing in the range of $10\%-90\%$ in steps of $10\%$, $L_{0.1},~L_{0.2},~...~,L_{0.9}$. The initial tidal radius, $\rt$ (green lines), and initial external field truncation radius, $\refe$(red lines), for the embedded star cluster are shown in the panels if they are $ < L_{0.95}$, where $ L_{0.95}$ is the Lagrangian radius enclosing $95\%$ mass of the system. For all the models with a SFE of $50\%$ and of $40\%$, the systems revirialised within a few tens of $\Myr$ after sudden gas expulsion. The $80\%$ Lagrangian radii, $L_{0.8}$, of all the models stay constant after the revirialisation, showing tiny oscillations with an amplitude of $\approx 1\%$. This implies that all the models with a SFE of $40\%-50\%$ can survive with $80\%$ of their initial mass. 
This is consistent with the results of bound mass fraction in Sec. \ref{mloss}. 

With a SFE of $30\%$, which is close to the critical value to leave a bound object in Newtonian dynamics, the $10\%$-$70\%$ Lagrangian radii for the most massive model, model $1$, remain stable after the expansion caused by the removal of gas, and the virial radius, $\refe$, cuts the model off at the radius of $\approx L_{0.7}$. For less massive models, models $2$-$4$, the evolution of all Lagrangian radii is very similar to the same models with larger values of the SFE. When the value of the SFE is reduced to $20\%$, the $\refe$ cuts model $1$ off at a radius slightly larger than the $L_{0.3}$ radius. The Lagrangian radii larger than the tidal radius keep expanding with time, which implies that stars outside the tidal radius are unbound and are escaping from the system. Interestingly, in the less massive models, models $2$-$4$, the $70\%$ Lagrangian radii remain stable after the revirialisation. When the value of SFE is further reduced, SFE$=10\%$, although $\rt>L_{0.1}$ for model $1$, $\refe$ is smaller than the $L_{0.1}$. The $10\%$ Lagrangian radii are not fully revirialised within $1~\Gyr$ and the fraction of bound mass is $\approx 6\%$ in the lower panel of Fig. \ref{fbound} for this model.  A larger fraction of bound mass for models $2$-$4$ is presented in Fig. \ref{lr}. The $10\%$-$30\%$ Lagrangian radii, $L_{0.1}-L_{0.3}$, are constants after $200~\Myr$. In model $3$, the $L_{0.4}$ is also stable. This is consistent with the results shown in Sec. \ref{mloss}.

The values of SFE are further decreased to $5\%$ and $2.5\%$, and all models are truncated by their tidal radii. The whole system of model $1$ expands with time and nothing is bound at the end of the simulations. The stable portion of the Lagrangian radii increases with declining initial mass of the models with the same SFE. However, due to the external field effect, the models are cut off at $\refe \ll  \rt$. The innermost $10\%$ Lagrangian radii for models $2$-$4$ with SFE of $5\%$ are stable in the late stage of the simulations. The evolution of Lagrangian radii with such a low SFE in MOND significantly differs from that in Newtonian dynamics. In the latter dynamics, any flat portion of Lagrangian radii cannot exist since nothing can be bound after the gas expulsion. Note that the smallest Lagrangian radii presented in Fig. \ref{lr} is the $L_{0.1}$ radius. Models $2$-$4$ with SFE=$2.5\%$ can survive with $2\%-3\%$ of their initial masses, but are not displayed in the bottom panels of Fig. \ref{lr} to avoid clustering.

\begin{figure}
\includegraphics[width=90mm]{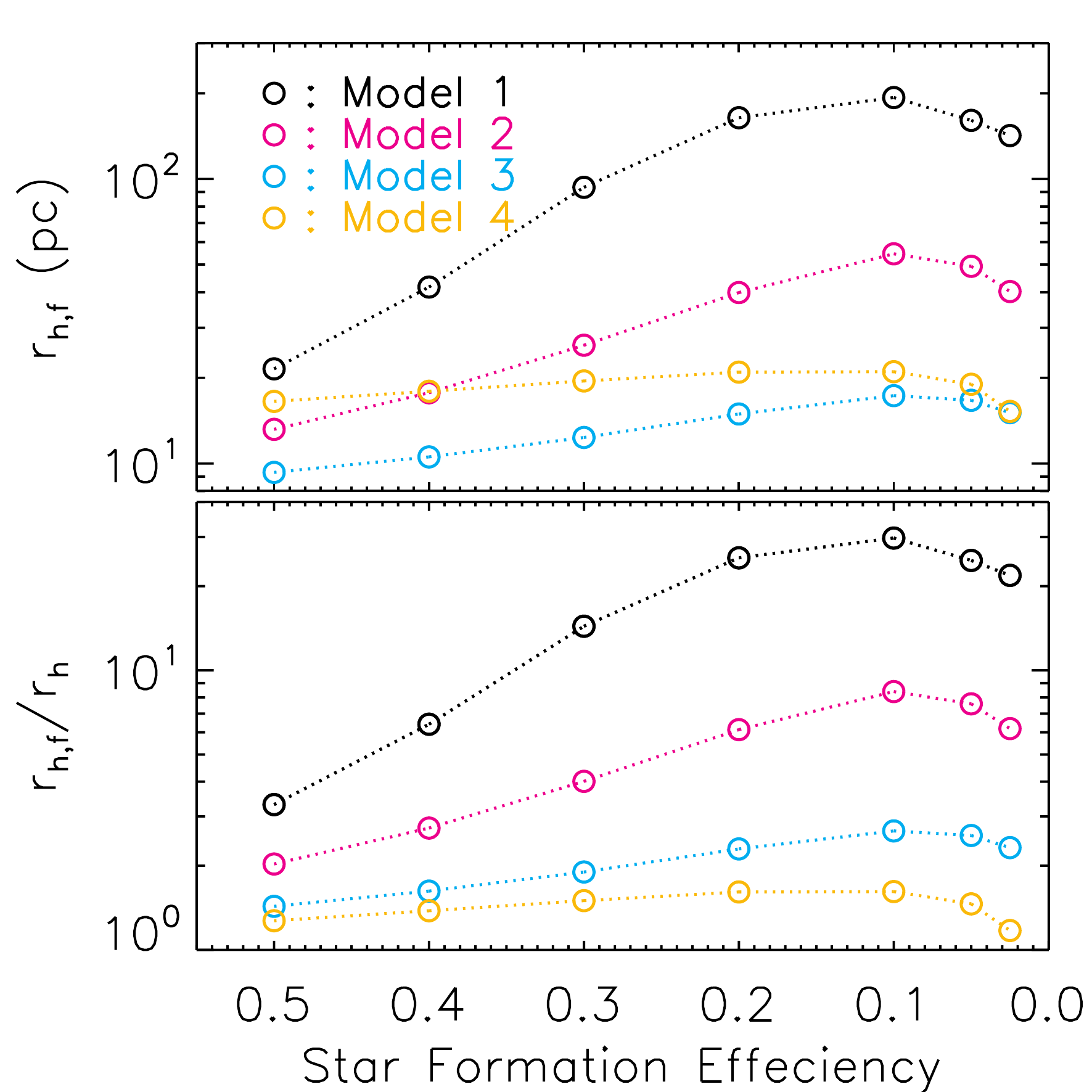}
\caption{Expansion after gas expulsion. Upper panel: $3$-dimensional half mass radius of the final remnants, $\rhf$. Lower panel: the ratios between $\rhf$ and $\rh$ of the embedded clusters listed in Table \ref{plummer}.}
\label{rh}
\end{figure}

Furthermore, there is a clear trend that for the same model, for example, model $2$, the Lagrangian radii at the end of the simulations are larger when the value of the SFE decreases. Such a trend indicates that the size of the remnant is larger with smaller SFE for the same initial density distribution of an embedded cluster. To quantify the expansion of the star clusters in MOND, the $3$-dimensional half mass radius of the final product (truncated at $\refe$), $\rhf$, and the expansion of size, i.e., the ratio between $\rhf$ and the initial $\rh$ of the embedded clusters, are displayed in Fig. \ref{rh}. The values of $\rhf$ grow with the decrease of the SFE in the range of $[10\%,~50\%]$ for all models, and approach the maximal value when SFE$=10\%$. $\rhf$ drops again when the SFE is smaller than $10\%$ for each model. \footnote{For model $1$ with a SFE of $5\%$ and $2.5\%$, the fraction of bound mass is almost zero (see Fig. \ref{fbound}). Therefore, the last two data points for model $1$ can be ignored.} The shapes of the $\rhf/\rh$ curves in Fig. \ref{rh} are very similar to that of $\rh$ for all the models.

In the most massive model, model $1$, $\rhf/\rh \approx 3.3$ for a SFE of $50\%$. This is a bit larger than that in Newtonian dynamics simulated by \citet{Baumgardt_Kroupa2007}, which is $2.95$ for an isolated star cluster. This should be attributed to the larger fraction of bound mass in MOND. $\rhf/\rh$ increases up to around $30$ for a SFE of $10\%$ for model $1$, which is $\approx 200~\pc$. This is a very diffuse remnant, with a bound mass of $\approx 6\times 10^4\msun$ (Table \ref{tab-fbound}). Such a stellar system is very similar to the ultra-faint and diffuse (UFD) Milky Way satellites, such as Ursa Major II, Leo T, Canes Venatici II and etc \citep{Simon_Geha2007}. For the less massive models, models $2$-$4$, $\rhf/\rh$ are much smaller than in model $1$. With a SFE of $50\%$, $\rhf/rh \in [1.3,~2.0]$ for models $2-4$, which is smaller than for the isolated model in \citet{Baumgardt_Kroupa2007}.
The more diffuse the initial model is, the smaller the value of $\rhf/\rh$ is with the same SFE. For the most diffuse model $4$, the value of $\rhf/\rh$ increases from $1.3$ to $1.6$ when the SFE decreases from $50\%$ to $10\%$, and then it declines to $1.2$ with a SFE$=2.5\%$. The size of the bound remnant does not increase as much as that in model $1$, given that model $4$ is in the deep MOND limit. In addition, with a SFE of $30\%$, which is often observed in embedded clusters and is very close to the canonical value for cluster survival of gas expulsion in Newtonian simulations, the value of $\rhf$ for model $4$ is about $20~\pc$ (see Fig. \ref{rh}), while the bound mass is about $2.7\times 10^4\msun$. The stellar mass and size of the final star cluster is very similar to the ultra-faint Milky Way satellites like Willman I and Segue I \citep{Simon_Geha2007}. There are several UFD satellite galaxies with half-mass radii larger than $200~\pc$, including Ursa Major I, Canes Venatici I and Hercules. But we also note that these satellites are located at larger Galactocentric distances than $100~\kpc$. As a result, both the virial radii and bound masses are larger in MOND. Therefore, MOND naturally provides a possible understanding for the formation of the UFD satellite galaxies without dark matter. 

To summarise, for an initially compact model, the size expansion is larger in MOND simulations owing to a larger fraction of bound mass, while for an initially diffuse model, the size expansion is smaller due to the much deeper MOND potential. The influence of sudden removal of gas on the size of the models is more significant in the Newtonian limit than that in the deep MOND limit, because in MOND the process is less destructive. In addition, only in the mild MOND gravity (i.e., model $1$) can very large expansion factor ($>20$) be reached.

\subsection{Mass density profiles}

\begin{figure*}
  \begin{centering}
\includegraphics[width=150mm]{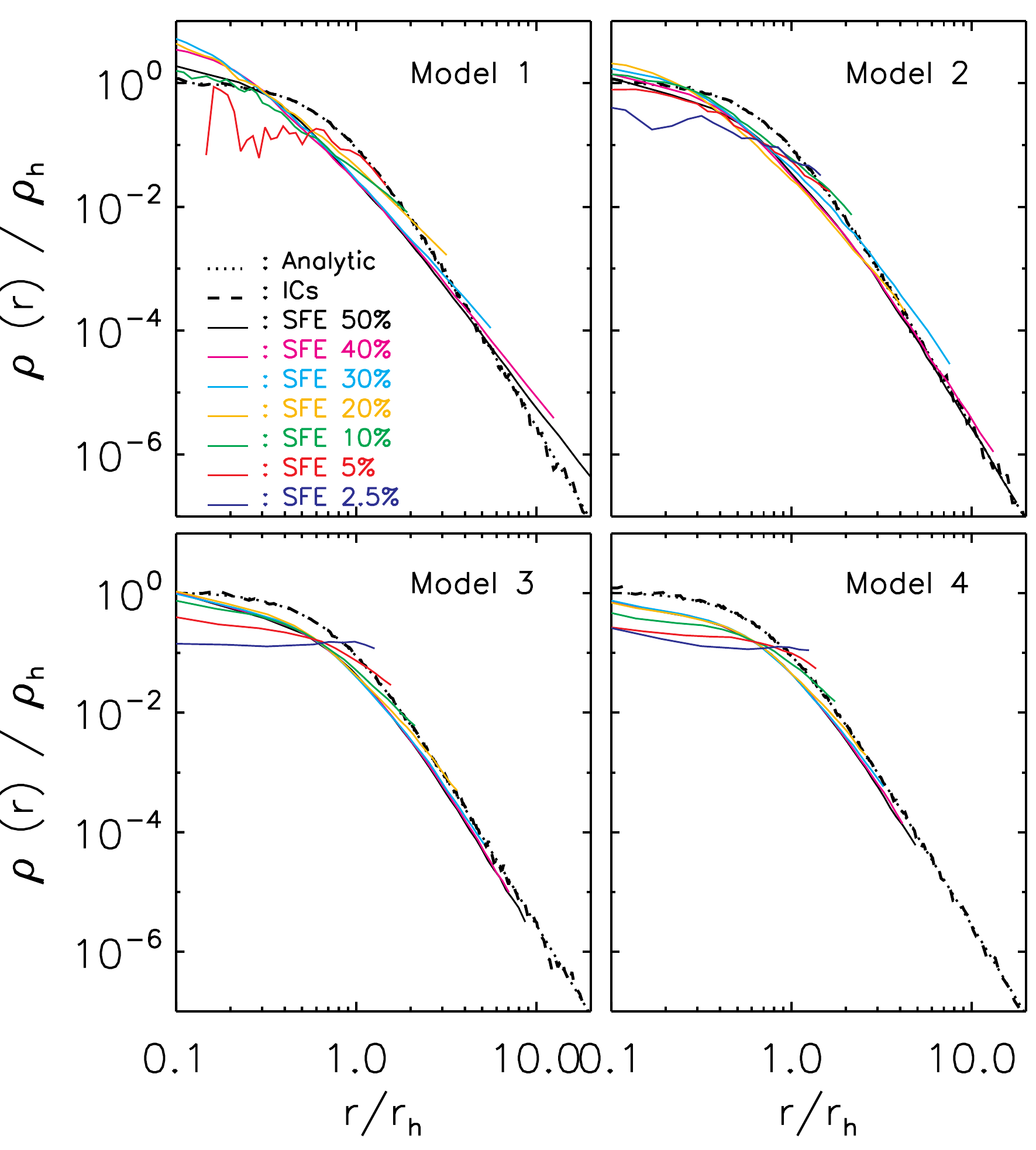}
\caption{The normalised density profiles of the four models. Different colours indicate the bound star clusters surviving gas expulsion with different values of the SFE. The dotted curves are the analytical Plummer density profile and the black dashed curves represent the density of the initial star clusters (ICs). The radius is normalised by $\rhf$ for the final bound systems and by $\rh$ for the ICs. The density is normalised by the average density, $\rho_{\rm h}$, within $\rhf$ ($\rh$ for the ICs). The models are truncated at $r=\refe$. }
\label{den}
\end{centering}
\end{figure*}

Fig. \ref{den} presents the mass density profiles of the initial embedded star clusters (the black dashed curves) and the bound star clusters after removing the gasous component (solid coloured curves). The horizontal axes are normalised by the $3$-dimensional half mass radius, $\rh$, for the initial star clusters (ICs) and $\rhf$ for the final products. The mass density is normalised by the mean density within $\rhf$ (or $\rh$ for the ICs),
\beq
\rho_{h}\equiv \frac{0.5M_*}{\rhf^3} = 0.5M \times {\rm SFE}\times f_{\rm bound}\rhf^{-3},
\eeq
where $M_*$ is the bound mass of the star clusters after sudden gas expulsion. Since the model star clusters are observed individually and each star cluster has its own density profile, the normalised density profiles of the models make it possible to inter-compare the concentration and slope of density profiles.

In model $1$, comparing to the ICs, the mass density profiles of the bound remnants are more concentrated within $0.3\rhf$ after gas expulsion with a SFE $\in [10\%,~50\%]$. The slopes of the density profiles fall faster than that of the ICs. This is very different to that of the clusters surviving in Newtonian dynamics. In fig. 2 of \citet{Boily_Kroupa2003}, the final profiles of the clusters with a SFE of $45\%-75\%$ in Newtonian dynamics still follow Plummer's mass distribution, i.e., the density distribution of the final products in the central region is flat. Here, however, model $1$ with a SFE of $30\%$ has the steepest density profile. The $\rho(r)$ profile for the remnant with a SFE of $5\%$ is much less concentrated compared to that of the ICs and appears rather flat within $\rhf$ with (particle Poisson) noise. Note that Model $1$ with a SFE of $2.5\%$ does not leave a bound core at the end of the simulation. In models $2-3$, the density profiles for the bound star clusters are also steeper than the ICs, but not as steep as that in model $1$. Moreover, the density profiles of the star clusters surviving the gas expulsion are less concentrated compared to that of the ICs when the SFE $=2.5\%$, which is similar to the case of model $1$ with a SFE of $5\%$. In other words, a core with constant density has been left in these models. For the most diffuse embedded cluster model, model $4$, the density profiles of the final products with all SFEs are less concentrated than those of the ICs. The $\rho(r)$ profiles of the bound remnants slowly decline with the radii in the centres where $r<\rhf$, and drop rapidly in the outer regions ($r>\rhf$) with a slope similar to that of the ICs.

To summarise, after gas expulsion in MOND, an originally massive embedded cluster leaves a bound star cluster with a more cuspy central density profile, while an initially less massive and more diffuse embedded cluster leaves a bound star cluster with a flat density profile within about its half-mass radius.

\section{Discussion and Conclusions}\label{summary}
In this work, we presented the first simulations of star clusters undergoing gas expulsion in Milgromian gravitation. A series of simulations with the SFE ranging from $2.5\%$ to $50\%$ are performed for embedded clusters with initial masses from $10^5\msun$ to $10^7\msun$. The fractions of bound masses, the Lagrangian radii, the half-mass radii and the mass density profiles of the surviving star clusters, are studied. We summarise and discuss our main results here. The kinematics (velocity dispersion profiles and velocity anisotropy profiles) will be presented in a subsequent paper.

The tidal radius for a system in MOND is much larger than that in Newtonian dynamics, while the uniform background gravitational field has a much stronger effect. Consequently, the star clusters are truncated at the virial radius where the strength of the external and internal fields are comparable \citep{Wu_Kroupa2015}.

For a given SFE,  after gas expulsion, the fraction of bound mass is larger in the deep MOND limit than in quasi-Newtonian gravity (mild MOND). In general, the star clusters can survive a low value of the SFE, $10\%$, for all the models in MOND, which is impossible if the gas expulsion is applied to the models in Newtonian dynamics. Furthermore, the initially deep MOND models with a critical SFE of $5\%$ and $2.5\%$ can leave a bound core. Within the framework of Newtonian dynamics, in order to leave a bound star cluster, the SFE should be at least $33\%$ in sudden gas expulsion \citep{Baumgardt_Kroupa2007}, which is apparently much larger than some observed SFEs in the dense cloud clumps of GMCs \citep[e.g., ][]{Megeath+2016}. By introducing additional physical processes in Newtonian gravitation, such as gradual gas expulsion, star clusters forming in complexes or initially non-equilibrium protoclusters, the SFE can be reduced to $15\%$ to leave a bound object\citep{Baumgardt_Kroupa2007,Fellhauer_Kroupa2005,Goodwin2009,Smith+2011,Shukirgaliyev+2017}. What is more, these additional physical process can be easily incorporated in MOND, which would further reduce the critical SFE.

The ultra-low SFEs allowed in MOND to yield bound stellar systems are relevant for the formation of the distant UFD satellites in the outer regions of the Milky Way, such as Hercules and Leo IV \citep{Geha+2013}. Moreover, the formation of ultra-faint tidal dwarf galaxies as seen in the Tadpole galaxy \citep{Kroupa2015}, could be another application of the ultra-low SFE in low density molecular clouds. 

The MOND computations show that a larger fraction of mass is bound to a surviving star cluster when the initial model for the embedded cluster is less massive. It implies that with a fixed value of the SFE, more stars are bound to the surviving star cluster when the embedded cluster model is in the deep MOND limit. The studies of the Lagrangian radii and the half-mass radii show that a more diffuse model expands less after the removal of gas. It implies that a mild MOND (quasi-Newtonian) system has a more substantial influence on the size of the surviving remnants than in the deep MOND limit. For a given SFE, the increase of size for a deep MOND system after sudden gas expulsion is much smaller than that for a quasi-Newtonian system. The mass density profiles of the surviving star clusters are more cuspy in the centre for an originally massive embedded cluster model dominated by quasi-Newtonian gravity, while the central density profile is flat for an originally less massive and more diffuse model dominated by MOND gravity.


Finally, since the potential of the bound GCs in MOND are significantly deeper to that in Newtonian dynamics, the kinamtics in the final GCs should be very different in the two dynamics.
We shall present an analysis of the kinematics in the two dynamics in a follow-up project.

\section{Acknowledgments}
The authors thank Luca Ciotti, Pasquale Londrillo and Carlo Nipoti for sharing their NMODY code. The authors acknowledge Ortwin Gerhard for sharing the code for generating the self-consistent N-body ICs using Lucy's method. XW thanks for support through NSFC grants 11503025, 11421303, Anhui NSF grant 1708085MA20, and ``the Fundamental Research Funds for the Central Universities''. XW thanks for support from ``Hundred Talents Project of Anhui Province''. This project was partially started when XW was an Alexander von Humboldt Fellow at the University of Bonn.

\bibliographystyle{aasjournal}
\bibliography{gasexp}

\begin{thebibliography}{}
\expandafter\ifx\csname natexlab\endcsname\relax\def\natexlab#1{#1}\fi
\providecommand{\url}[1]{\href{#1}{#1}}

\bibitem[{{Angus}(2009)}]{Angus2009}
{Angus}, G.~W. 2009, \mnras, 394, 527

\bibitem[{{Angus} \& {Diaferio}(2011)}]{Angus_Diaferio2011}
{Angus}, G.~W., \& {Diaferio}, A. 2011, \mnras, 417, 941

\bibitem[{{Angus} {et~al.}(2014){Angus}, {Diaferio}, {Famaey}, {Gentile}, \&
  {van der Heyden}}]{Angus+2014b}
{Angus}, G.~W., {Diaferio}, A., {Famaey}, B., {Gentile}, G., \& {van der Heyden}, K.~J. 2014, JCAP, 10, 079

\bibitem[{{Angus} {et~al.}(2013){Angus}, {Diaferio}, {Famaey}, \& {van der
  Heyden}}]{Angus+2013}
{Angus}, G.~W., {Diaferio}, A., {Famaey}, B., \& {van der Heyden}, K.~J. 2013,
  \mnras, 436, 202

\bibitem[{{Angus} {et~al.}(2007){Angus}, {Shan}, {Zhao}, \&
  {Famaey}}]{Angus+2007}
{Angus}, G.~W., {Shan}, H.~Y., {Zhao}, H.~S., \& {Famaey}, B. 2007, \apjl, 654,
  L13

\bibitem[{{Banerjee} \& {Kroupa}(2012)}]{Banerjee_Kroupa2012}
{Banerjee}, S., \& {Kroupa}, P. 2012, \aap, 547, A23

\bibitem[{{Banerjee} \& {Kroupa}(2015)}]{Banerjee_Kroupa2015}
---. 2015, \mnras, 447, 728

\bibitem[{{Banerjee} \& {Kroupa}(2017)}]{Banerjee_Kroupa2017}
---. 2017, \aap, 597, A28

\bibitem[{{Banik} \& {Zhao}(2018)}]{Banik_Zhao2018}
{Banik}, I., \& {Zhao}, H. 2018, \mnras, 473, 419

\bibitem[{{Bastian} \& {Goodwin}(2006)}]{Bastian_Goodwin2006}
{Bastian}, N., \& {Goodwin}, S.~P. 2006, \mnras, 369, L9

\bibitem[{{Baumgardt} {et~al.}(2009){Baumgardt}, {C{\^o}t{\'e}}, {Hilker},
  {Rejkuba}, {Mieske}, {Djorgovski}, \& {Stetson}}]{Baumgardt+2009}
{Baumgardt}, H., {C{\^o}t{\'e}}, P., {Hilker}, M., {et~al.} 2009, \mnras, 396,
  2051

\bibitem[{{Baumgardt} {et~al.}(2005){Baumgardt}, {Grebel}, \&
  {Kroupa}}]{Baumgardt+2005}
{Baumgardt}, H., {Grebel}, E.~K., \& {Kroupa}, P. 2005, \mnras, 359, L1

\bibitem[{{Baumgardt} \& {Kroupa}(2007)}]{Baumgardt_Kroupa2007}
{Baumgardt}, H., \& {Kroupa}, P. 2007, \mnras, 380, 1589

\bibitem[{{Bekenstein} \& {Milgrom}(1984)}]{BM1984}
{Bekenstein}, J., \& {Milgrom}, M. 1984, \apj, 286, 7

\bibitem[{{Bekenstein}(2004)}]{Bekenstein2004}
{Bekenstein}, J.~D. 2004, \prd, 70, 083509

\bibitem[{{Bekki} {et~al.}(2017){Bekki}, {Je{\v r}{\'a}bkov{\'a}}, \&
  {Kroupa}}]{Bekki+2017}
{Bekki}, K., {Je{\v r}{\'a}bkov{\'a}}, T., \& {Kroupa}, P. 2017, \mnras, 471,
  2242

\bibitem[{{Belloni} {et~al.}(2017){Belloni}, {Kroupa}, {Rocha-Pinto}, \&
  {Giersz}}]{Belloni+2017}
{Belloni}, D., {Kroupa}, P., {Rocha-Pinto}, H.~J., \& {Giersz}, M. 2017, ArXiv
  e-prints, arXiv:1711.07987

\bibitem[{{B{\'{\i}}lek} {et~al.}(2014){B{\'{\i}}lek}, {Barto{\v s}kov{\'a}},
  {Ebrov{\'a}}, \& {Jungwiert}}]{Bilek+2014}
{B{\'{\i}}lek}, M., {Barto{\v s}kov{\'a}}, K., {Ebrov{\'a}}, I., \&
  {Jungwiert}, B. 2014, \aap, 566, A151

\bibitem[{{B{\'{\i}}lek} {et~al.}(2013){B{\'{\i}}lek}, {Jungwiert},
  {J{\'{\i}}lkov{\'a}}, {Ebrov{\'a}}, {Barto{\v s}kov{\'a}}, \& {K{\v
  r}{\'{\i}}{\v z}ek}}]{Bilek+2013}
{B{\'{\i}}lek}, M., {Jungwiert}, B., {J{\'{\i}}lkov{\'a}}, L., {et~al.} 2013,
  \aap, 559, A110

\bibitem[{{Boily} \& {Kroupa}(2003)}]{Boily_Kroupa2003}
{Boily}, C.~M., \& {Kroupa}, P. 2003, \mnras, 338, 673

\bibitem[{{Brinkmann} {et~al.}(2017){Brinkmann}, {Banerjee}, {Motwani}, \&
  {Kroupa}}]{Brinkmann+2017}
{Brinkmann}, N., {Banerjee}, S., {Motwani}, B., \& {Kroupa}, P. 2017, \aap,
  600, A49

\bibitem[{{Clowe} {et~al.}(2006){Clowe}, {Brada{\v c}}, {Gonzalez},
  {Markevitch}, {Randall}, {Jones}, \& {Zaritsky}}]{Clowe+2006}
{Clowe}, D., {Brada{\v c}}, M., {Gonzalez}, A.~H., {et~al.} 2006, \apjl, 648,
  L109

\bibitem[{{Dib} {et~al.}(2013){Dib}, {Gutkin}, {Brandner}, \&
  {Basu}}]{Dib+2013}
{Dib}, S., {Gutkin}, J., {Brandner}, W., \& {Basu}, S. 2013, \mnras, 436, 3727

\bibitem[{{Duarte-Cabral} {et~al.}(2013){Duarte-Cabral}, {Bontemps}, {Motte},
  {Hennemann}, {Schneider}, \& {Andr{\'e}}}]{Duarte-Cabral+2013}
{Duarte-Cabral}, A., {Bontemps}, S., {Motte}, F., {et~al.} 2013, \aap, 558,
  A125

\bibitem[{{Durazo} {et~al.}(2017){Durazo}, {Hernandez}, {Cervantes Sodi}, \&
  {S{\'a}nchez}}]{Durazo+2017}
{Durazo}, R., {Hernandez}, X., {Cervantes Sodi}, B., \& {S{\'a}nchez}, S.~F.
  2017, \apj, 837, 179

\bibitem[{{Evans}(1991)}]{Evans1991}
{Evans}, II, N.~J. 1991, in Astronomical Society of the Pacific Conference
  Series, Vol.~20, Frontiers of Stellar Evolution, ed. D.~L. {Lambert}, 45--95

\bibitem[{{Famaey} \& {Binney}(2005)}]{Famaey_Binney2005}
{Famaey}, B., \& {Binney}, J. 2005, \mnras, 363, 603

\bibitem[{{Famaey} {et~al.}(2007){Famaey}, {Bruneton}, \& {Zhao}}]{Famaey+2007}
{Famaey}, B., {Bruneton}, J.-P., \& {Zhao}, H. 2007, \mnras, 377, L79

\bibitem[{{Famaey} \& {McGaugh}(2012)}]{Famaey_McGaugh2012}
{Famaey}, B., \& {McGaugh}, S.~S. 2012, Living Reviews in Relativity, 15, 10

\bibitem[{{Fellhauer} \& {Kroupa}(2005)}]{Fellhauer_Kroupa2005}
{Fellhauer}, M., \& {Kroupa}, P. 2005, \apj, 630, 879

\bibitem[{{Frank} {et~al.}(2012){Frank}, {Hilker}, {Baumgardt}, {C{\^o}t{\'e}},
  {Grebel}, {Haghi}, {K{\"u}pper}, \& {Djorgovski}}]{Frank+2012}
{Frank}, M.~J., {Hilker}, M., {Baumgardt}, H., {et~al.} 2012, \mnras, 423, 2917

\bibitem[{{Geha} {et~al.}(2013){Geha}, {Brown}, {Tumlinson}, {Kalirai},
  {Simon}, {Kirby}, {VandenBerg}, {Mu{\~n}oz}, {Avila}, {Guhathakurta}, \&
  {Ferguson}}]{Geha+2013}
{Geha}, M., {Brown}, T.~M., {Tumlinson}, J., {et~al.} 2013, \apj, 771, 29

\bibitem[{{Gentile} {et~al.}(2010){Gentile}, {Famaey}, {Angus}, \&
  {Kroupa}}]{Gentile+2010}
{Gentile}, G., {Famaey}, B., {Angus}, G., \& {Kroupa}, P. 2010, \aap, 509, A97

\bibitem[{{Gerhard}(1991)}]{Gerhard1991}
{Gerhard}, O.~E. 1991, \mnras, 250, 812

\bibitem[{{Gieles} {et~al.}(2012){Gieles}, {Moeckel}, \&
  {Clarke}}]{Gieles+2012}
{Gieles}, M., {Moeckel}, N., \& {Clarke}, C.~J. 2012, \mnras, 426, L11

\bibitem[{{Goodwin}(1997)}]{Goodwin1997}
{Goodwin}, S.~P. 1997, \mnras, 284, 785

\bibitem[{{Goodwin}(2009)}]{Goodwin2009}
---. 2009, \apss, 324, 259

\bibitem[{{Goodwin} \& {Bastian}(2006)}]{Goodwin_Bastian2006}
{Goodwin}, S.~P., \& {Bastian}, N. 2006, \mnras, 373, 752

\bibitem[{{Harris}(1996)}]{Harris1996}
{Harris}, W.~E. 1996, \aj, 112, 1487

\bibitem[{{Hernandez} \& {Jim{\'e}nez}(2012)}]{Hernandez_Jimenez2012}
{Hernandez}, X., \& {Jim{\'e}nez}, M.~A. 2012, \apj, 750, 9

\bibitem[{{Hernandez} {et~al.}(2013){Hernandez}, {Jim{\'e}nez}, \&
  {Allen}}]{Hernandez+2013}
{Hernandez}, X., {Jim{\'e}nez}, M.~A., \& {Allen}, C. 2013, \mnras, 428, 3196

\bibitem[{{Higuchi} {et~al.}(2009){Higuchi}, {Kurono}, {Saito}, \&
  {Kawabe}}]{Higuchi+2009}
{Higuchi}, A.~E., {Kurono}, Y., {Saito}, M., \& {Kawabe}, R. 2009, \apj, 705,
  468

\bibitem[{{Hopkins} {et~al.}(2013){Hopkins}, {Narayanan}, {Murray}, \&
  {Quataert}}]{Hopkins+2013}
{Hopkins}, P.~F., {Narayanan}, D., {Murray}, N., \& {Quataert}, E. 2013,
  \mnras, 433, 69

\bibitem[{{Ibata} {et~al.}(2011{\natexlab{a}}){Ibata}, {Sollima}, {Nipoti},
  {Bellazzini}, {Chapman}, \& {Dalessandro}}]{Ibata+2011a}
{Ibata}, R., {Sollima}, A., {Nipoti}, C., {et~al.} 2011{\natexlab{a}}, \apj,
  738, 186

\bibitem[{{Ibata} {et~al.}(2011{\natexlab{b}}){Ibata}, {Sollima}, {Nipoti},
  {Bellazzini}, {Chapman}, \& {Dalessandro}}]{Ibata+2011b}
---. 2011{\natexlab{b}}, \apj, 743, 43

\bibitem[{{Jee} {et~al.}(2012){Jee}, {Mahdavi}, {Hoekstra}, {Babul},
  {Dalcanton}, {Carroll}, \& {Capak}}]{Jee+2012}
{Jee}, M.~J., {Mahdavi}, A., {Hoekstra}, H., {et~al.} 2012, \apj, 747, 96

\bibitem[{{Jordi} {et~al.}(2009){Jordi}, {Grebel}, {Hilker}, {Baumgardt},
  {Frank}, {Kroupa}, {Haghi}, {C{\^o}t{\'e}}, \& {Djorgovski}}]{Jordi+2009}
{Jordi}, K., {Grebel}, E.~K., {Hilker}, M., {et~al.} 2009, \aj, 137, 4586

\bibitem[{{Kainulainen} {et~al.}(2014){Kainulainen}, {Federrath}, \&
  {Henning}}]{Kainulainen+2014}
{Kainulainen}, J., {Federrath}, C., \& {Henning}, T. 2014, Science, 344, 183

\bibitem[{{Kennicutt}(1998)}]{Kennicutt1998}
{Kennicutt}, Jr., R.~C. 1998, \apj, 498, 541

\bibitem[{{Kroupa}(1995)}]{Kroupa1995a}
{Kroupa}, P. 1995, \mnras, 277, astro-ph/9508117

\bibitem[{{Kroupa}(2012)}]{Kroupa2012}
---. 2012, ArXiv e-prints, arXiv:1204.2546

\bibitem[{{Kroupa}(2015)}]{Kroupa2015}
---. 2015, Canadian Journal of Physics, 93, 169

\bibitem[{{Kroupa} {et~al.}(2001){Kroupa}, {Aarseth}, \&
  {Hurley}}]{Kroupa+2001}
{Kroupa}, P., {Aarseth}, S., \& {Hurley}, J. 2001, \mnras, 321, 699

\bibitem[{{Kroupa} {et~al.}(2012){Kroupa}, {Pawlowski}, \&
  {Milgrom}}]{Kroupa+2012}
{Kroupa}, P., {Pawlowski}, M., \& {Milgrom}, M. 2012, International Journal of
  Modern Physics D, 21, 1230003

\bibitem[{{Krumholz} \& {Matzner}(2009)}]{Krumholz_Matzner2009}
{Krumholz}, M.~R., \& {Matzner}, C.~D. 2009, \apj, 703, 1352

\bibitem[{{Lada}(2010)}]{Lada2010}
{Lada}, C.~J. 2010, Philosophical Transactions of the Royal Society of London
  Series A, 368, 713

\bibitem[{{Lada} \& {Lada}(2003)}]{Lada_Lada2003}
{Lada}, C.~J., \& {Lada}, E.~A. 2003, \araa, 41, 57

\bibitem[{{Lada} {et~al.}(1984){Lada}, {Margulis}, \& {Dearborn}}]{Lada+1984}
{Lada}, C.~J., {Margulis}, M., \& {Dearborn}, D. 1984, \apj, 285, 141

\bibitem[{{Lane} {et~al.}(2009){Lane}, {Kiss}, {Lewis}, {Ibata}, {Siebert},
  {Bedding}, \& {Sz{\'e}kely}}]{Lane+2009}
{Lane}, R.~R., {Kiss}, L.~L., {Lewis}, G.~F., {et~al.} 2009, \mnras, 400, 917

\bibitem[{{Lane} {et~al.}(2010){Lane}, {Kiss}, {Lewis}, {Ibata}, {Siebert},
  {Bedding}, {Sz{\'e}kely}, {Balog}, \& {Szab{\'o}}}]{Lane+2010}
---. 2010, \mnras, 406, 2732

\bibitem[{{Londrillo} \& {Nipoti}(2009)}]{nmody}
{Londrillo}, P., \& {Nipoti}, C. 2009, Memorie della Societa Astronomica
  Italiana Supplement, 13, 89

\bibitem[{{Lucy}(1974)}]{Lucy1974}
{Lucy}, L.~B. 1974, \aj, 79, 745

\bibitem[{{L{\"u}ghausen} {et~al.}(2013){L{\"u}ghausen}, {Famaey}, {Kroupa},
  {Angus}, {Combes}, {Gentile}, {Tiret}, \& {Zhao}}]{Lughausen+2013}
{L{\"u}ghausen}, F., {Famaey}, B., {Kroupa}, P., {et~al.} 2013, \mnras, 432,
  2846

\bibitem[{{Marks} {et~al.}(2012){Marks}, {Kroupa}, {Dabringhausen}, \&
  {Pawlowski}}]{Marks+2012}
{Marks}, M., {Kroupa}, P., {Dabringhausen}, J., \& {Pawlowski}, M.~S. 2012,
  \mnras, 422, 2246

\bibitem[{{McGaugh} \& {Milgrom}(2013)}]{McGaugh_Milgrom2013}
{McGaugh}, S., \& {Milgrom}, M. 2013, \apj, 775, 139

\bibitem[{{McGaugh}(2004)}]{McGaugh2004}
{McGaugh}, S.~S. 2004, \apj, 609, 652

\bibitem[{{McGaugh}(2016)}]{McGaugh2016}
---. 2016, \apjl, 832, L8

\bibitem[{{Megeath} {et~al.}(2016){Megeath}, {Gutermuth}, {Muzerolle},
  {Kryukova}, {Hora}, {Allen}, {Flaherty}, {Hartmann}, {Myers}, {Pipher},
  {Stauffer}, {Young}, \& {Fazio}}]{Megeath+2016}
{Megeath}, S.~T., {Gutermuth}, R., {Muzerolle}, J., {et~al.} 2016, \aj, 151, 5

\bibitem[{{Milgrom}(1983{\natexlab{a}})}]{Milgrom1983a}
{Milgrom}, M. 1983{\natexlab{a}}, \apj, 270, 365

\bibitem[{{Milgrom}(1983{\natexlab{b}})}]{Milgrom1983b}
---. 1983{\natexlab{b}}, \apj, 270, 371

\bibitem[{{Milgrom}(1994)}]{Milgrom1994}
---. 1994, \apj, 429, 540

\bibitem[{{Milgrom}(2009{\natexlab{a}})}]{Milgrom2009b}
---. 2009{\natexlab{a}}, \prd, 80, 123536

\bibitem[{{Milgrom}(2009{\natexlab{b}})}]{Milgrom2009}
---. 2009{\natexlab{b}}, \mnras, 398, 1023

\bibitem[{{Milgrom}(2010)}]{Milgrom2010}
---. 2010, \mnras, 405, 1129

\bibitem[{{Milgrom}(2016)}]{Milgrom2016}
---. 2016, Physical Review Letters, 117, 141101

\bibitem[{{Moore}(1996)}]{Moore1996}
{Moore}, B. 1996, \apjl, 461, L13

\bibitem[{{Motte} {et~al.}(2017){Motte}, {Bontemps}, \& {Louvet}}]{Motte+2017}
{Motte}, F., {Bontemps}, S., \& {Louvet}, F. 2017, ArXiv e-prints,
  arXiv:1706.00118

\bibitem[{{Murray}(2011)}]{Murray2011}
{Murray}, N. 2011, \apj, 729, 133

\bibitem[{{Murray} {et~al.}(2010){Murray}, {Quataert}, \&
  {Thompson}}]{Murray+2010}
{Murray}, N., {Quataert}, E., \& {Thompson}, T.~A. 2010, \apj, 709, 191

\bibitem[{{Natarajan} \& {Zhao}(2008)}]{Natarajan_Zhao2008}
{Natarajan}, P., \& {Zhao}, H. 2008, \mnras, 389, 250

\bibitem[{{Nipoti} {et~al.}(2008){Nipoti}, {Ciotti}, {Binney}, \&
  {Londrillo}}]{Nipoti+2008}
{Nipoti}, C., {Ciotti}, L., {Binney}, J., \& {Londrillo}, P. 2008, \mnras, 386,
  2194

\bibitem[{{Nipoti} {et~al.}(2011){Nipoti}, {Ciotti}, \&
  {Londrillo}}]{Nipoti+2011}
{Nipoti}, C., {Ciotti}, L., \& {Londrillo}, P. 2011, \mnras, 414, 3298

\bibitem[{{Nipoti} {et~al.}(2007){Nipoti}, {Londrillo}, \&
  {Ciotti}}]{Nipoti+2007}
{Nipoti}, C., {Londrillo}, P., \& {Ciotti}, L. 2007, \apj, 660, 256

\bibitem[{{Pfalzner} \& {Kaczmarek}(2013)}]{Pfalzner_Kaczmarek2013}
{Pfalzner}, S., \& {Kaczmarek}, T. 2013, \aap, 555, A135

\bibitem[{{Plummer}(1911)}]{Plummer1911}
{Plummer}, H.~C. 1911, \mnras, 71, 460

\bibitem[{{Sanders}(1990)}]{Sanders1990}
{Sanders}, R.~H. 1990, \aapr, 2, 1

\bibitem[{{Sanders}(2012{\natexlab{a}})}]{Sanders2012a}
---. 2012{\natexlab{a}}, \mnras, 419, L6

\bibitem[{{Sanders}(2012{\natexlab{b}})}]{Sanders2012b}
---. 2012{\natexlab{b}}, \mnras, 422, L21

\bibitem[{{Sanders} \& {McGaugh}(2002)}]{Sanders_McGaugh2002}
{Sanders}, R.~H., \& {McGaugh}, S.~S. 2002, \araa, 40, 263

\bibitem[{{Scarpa}(2006)}]{Scarpa2006}
{Scarpa}, R. 2006, in American Institute of Physics Conference Series, Vol.
  822, First Crisis in Cosmology Conference, ed. E.~J. {Lerner} \& J.~B.
  {Almeida}, 253--265

\bibitem[{{Scarpa} \& {Falomo}(2010)}]{Scarpa_Falomo2010}
{Scarpa}, R., \& {Falomo}, R. 2010, \aap, 523, A43

\bibitem[{{Scarpa} {et~al.}(2011){Scarpa}, {Marconi}, {Carraro}, {Falomo}, \&
  {Villanova}}]{Scarpa+2011}
{Scarpa}, R., {Marconi}, G., {Carraro}, G., {Falomo}, R., \& {Villanova}, S.
  2011, \aap, 525, A148

\bibitem[{{Scarpa} {et~al.}(2007){Scarpa}, {Marconi}, {Gilmozzi}, \&
  {Carraro}}]{Scarpa+2007}
{Scarpa}, R., {Marconi}, G., {Gilmozzi}, R., \& {Carraro}, G. 2007, \aap, 462,
  L9

\bibitem[{{Shukirgaliyev} {et~al.}(2017){Shukirgaliyev}, {Parmentier},
  {Berczik}, \& {Just}}]{Shukirgaliyev+2017}
{Shukirgaliyev}, B., {Parmentier}, G., {Berczik}, P., \& {Just}, A. 2017, \aap,
  605, A119

\bibitem[{{Simon} \& {Geha}(2007)}]{Simon_Geha2007}
{Simon}, J.~D., \& {Geha}, M. 2007, \apj, 670, 313

\bibitem[{{Skordis} {et~al.}(2006){Skordis}, {Mota}, {Ferreira}, \&
  {B{\oe}hm}}]{Skordis+2006}
{Skordis}, C., {Mota}, D.~F., {Ferreira}, P.~G., \& {B{\oe}hm}, C. 2006,
  Physical Review Letters, 96, 011301

\bibitem[{{Smith} {et~al.}(2011){Smith}, {Fellhauer}, {Goodwin}, \&
  {Assmann}}]{Smith+2011}
{Smith}, R., {Fellhauer}, M., {Goodwin}, S., \& {Assmann}, P. 2011, \mnras,
  414, 3036

\bibitem[{{Thomas} {et~al.}(2017){Thomas}, {Famaey}, {Ibata}, {Renaud},
  {Martin}, \& {Kroupa}}]{Thomas+2017}
{Thomas}, G.~F., {Famaey}, B., {Ibata}, R., {et~al.} 2017, ArXiv e-prints,
  arXiv:1709.01934

\bibitem[{{Wu} {et~al.}(2008){Wu}, {Famaey}, {Gentile}, {Perets}, \&
  {Zhao}}]{Wu+2008}
{Wu}, X., {Famaey}, B., {Gentile}, G., {Perets}, H., \& {Zhao}, H. 2008,
  \mnras, 386, 2199

\bibitem[{{Wu} \& {Kroupa}(2013)}]{Wu+2013}
{Wu}, X., \& {Kroupa}, P. 2013, \mnras, 435, 728

\bibitem[{{Wu} \& {Kroupa}(2015)}]{Wu_Kroupa2015}
---. 2015, \mnras, 446, 330

\bibitem[{{Wu} {et~al.}(2017){Wu}, {Wang}, {Feix}, \& {Zhao}}]{Wu+2017}
{Wu}, X., {Wang}, Y., {Feix}, M., \& {Zhao}, H. 2017, \apj, 844, 130

\bibitem[{{Wu} {et~al.}(2010){Wu}, {Zhao}, \& {Famaey}}]{Wu+2010}
{Wu}, X., {Zhao}, H., \& {Famaey}, B. 2010, JCAP, 6, 010

\bibitem[{{Wu} {et~al.}(2007){Wu}, {Zhao}, {Famaey}, {Gentile}, {Tiret},
  {Combes}, {Angus}, \& {Robin}}]{Wu+2007}
{Wu}, X., {Zhao}, H., {Famaey}, B., {et~al.} 2007, \apjl, 665, L101

\bibitem[{{Wuchterl} \& {Tscharnuter}(2003)}]{Wuchterl_Tscharnuter2003}
{Wuchterl}, G., \& {Tscharnuter}, W.~M. 2003, \aap, 398, 1081

\bibitem[{{Zhao} \& {Tian}(2006)}]{Zhao_Tian2006}
{Zhao}, H., \& {Tian}, L. 2006, \aap, 450, 1005

\bibitem[{{Zhao}(2008)}]{Zhao2008}
{Zhao}, H.~S. 2008, in Journal of Physics Conference Series, Vol. 140, Journal
  of Physics Conference Series, 012002

\bibitem[{{Zhao} \& {Famaey}(2006)}]{Zhao_Famaey2006}
{Zhao}, H.~S., \& {Famaey}, B. 2006, \apjl, 638, L9

\end{thebibliography}
\end{document}